\documentclass[10pt,letterpaper]{article}
\usepackage[top=0.85in,left=2.75in,footskip=0.75in]{geometry}

\usepackage{amsmath,amssymb}

\usepackage{changepage}
\usepackage{tabularx}
\usepackage[table]{xcolor}
\usepackage{float}
\usepackage{booktabs}
\usepackage{textcomp,marvosym}

\usepackage{cite}

\usepackage{nameref,hyperref}

\usepackage[right]{lineno}

\usepackage[nopatch=eqnum]{microtype}
\DisableLigatures[f]{encoding = *, family = * }

\usepackage[table]{xcolor}

\usepackage{array}

\newcolumntype{+}{!{\vrule width 2pt}}

\newlength\savedwidth

\raggedright
\usepackage[aboveskip=1pt,labelfont=bf,labelsep=period,justification=raggedright,singlelinecheck=off]{caption}

\makeatletter
\renewcommand{\@biblabel}[1]{\quad#1.}
\makeatother

\usepackage{lastpage,fancyhdr,graphicx}
\usepackage{epstopdf}
\fancyheadoffset[L]{2.25in}
\fancyfootoffset[L]{2.25in}
\begin{document}
\vspace*{0.2in}

\begin{flushleft}
{\Large
\textbf\newline{Interpretable Machine Learning Model for Early Prediction of 30-Day Mortality in ICU Patients With Coexisting Hypertension and Atrial Fibrillation: A Retrospective Cohort Study} 
}
\newline
\\
Shuheng Chen\textsuperscript{1},
Yong Si\textsuperscript{1},
Junyi Fan\textsuperscript{1},
Li Sun\textsuperscript{1},
Greg Placencia\textsuperscript{2},
Elham Pishgar\textsuperscript{3},
Kamiar Alaei\textsuperscript{4},
Maryam Pishgar\textsuperscript{1,*}
\\
\bigskip
\textbf{1} Department ofIndustrial and Systems Engineering, University of Southern California, Los Angeles, California, United States
\\
\textbf{2} Department of Industrial and Manufacturing Engineering, California State Polytechnic University Pomona, Pomona, California, United States
\\
\textbf{3} Colorectal Research Center, lran University of Medical Sciences, Tehran, Iran
\\
\textbf{4} Department of Health Science, California State University Long Beach, Long Beach, California, United States
\\
\bigskip






* pishgar@usc.edu

\end{flushleft}
\section*{Abstract}
\textbf{Background:}
Hypertension and its common complication, atrial fibrillation (AF), frequently coexist in critically ill patients and are strongly associated with significantly elevated mortality rates in the intensive care unit (ICU). Early identification of high-risk individuals is crucial for targeted interventions and optimal resource allocation. However, limited research has focused on short-term mortality prediction within this complex patient subgroup.

\textbf{Methods:}
This study conducted a retrospective analysis of 1,301 adult ICU patients with concurrent hypertension and AF using the MIMIC-IV database. Structured data, including chart events, laboratory results, procedure events, drug administration, as well as demographic and comorbidity information from the first 24 hours of ICU admission, were extracted. Following quality control, missing data imputation, and feature selection, and under the guidance of clinical experts, 17 clinically interpretable variables were retained. The cohort was stratified and split into training (70\%) and test (30\%) sets. To address class imbalance, outcome-weighted training was applied. The CatBoost model, alongside five baseline machine learning models-LightGBM, XGBoost, logistic regression, Naive Bayes, and neural networks—was benchmarked using five-fold cross-validation, with the Area Under the Receiver Operating Characteristic Curve (AUROC) as the primary performance metric. Model interpretability was evaluated through an ablation study, SHapley Additive exPlanations (SHAP), Accumulated Local Effects (ALE), and DREAM analyses.

\textbf{Results:}
The CatBoost model demonstrated strong predictive performance with an AUROC of 0.889 (95\% CI: 0.840–0.924), accuracy of 0.831, F1-score of 0.522, sensitivity of 0.837, specificity of 0.830, PPV of 0.379, and NPV of 0.976. Global and local explanation methods, including SHAP, ALE, and DREAM analyses, identified key predictors such as the Richmond-RAS Scale, pO\textsubscript{2}, CefePIME, and Invasive Ventilation, highlighting the model's robustness and clinical applicability in predicting 30-day mortality in ICU patients with hypertension and AF.

\textbf{Conclusion:}
The proposed model demonstrates strong performance and high interpretability in early mortality prediction, enabling early intervention and personalized care decisions in this specific population. Future work will focus on external validation using multi-center ICU datasets, incorporating longitudinal data to monitor patient trajectories, and extending the approach to other diseases using the developed pipeline.

\section*{Author summary}
S.C. was responsible for the study design, development of the methodology, execution of the experiments, data analysis, and drafting the initial manuscript. J.F., Y.S., and L.S. contributed to the experimental work and participated in manuscript preparation. E.P., K.A., and G.P. provided insightful feedback on both the study design and manuscript. M.P. oversaw the project and offered comprehensive guidance. All authors reviewed and gave their approval for the final manuscript.


\section*{Introduction}
Hypertension, characterized by sustained elevated blood pressure, arises from a complex interplay of genetic predisposition, unhealthy dietary habits, obesity, physical inactivity, and chronic stress \cite{carey2018prevention,mills2020global,singh2023genome,mohammed2016dietary,shariq2020obesity}. As one of the most prevalent chronic conditions worldwide, hypertension affects over one billion adults, with its prevalence continuing to rise, particularly in low- and middle-income countries \cite{cureau2021worldwide,WHO2023,WHF2023}. In addition to the substantial health risks, hypertension places a considerable economic burden on healthcare systems globally, encompassing the costs of treatment, management of complications, and productivity losses \cite{wierzejska2020global,go2014effective,constant2016economic,gnugesser2022economic,sorato2022societal}. Due to its often asymptomatic nature in the early stages, hypertension frequently goes undiagnosed, leading to delays in initiating treatment. This delay results in significant contributions to severe cardiovascular diseases, including stroke, heart failure, myocardial infarction, chronic kidney disease, and atrial fibrillation (AF) \cite{messerli2007essential,kjeldsen2018hypertension,burnier2023hypertension}.

One of the most significant complications arising from uncontrolled hypertension is AF, and it is particularly concerning in intensive care unit (ICU) patients due to the elevated risk of adverse outcomes, including increased mortality rates \cite{antoun2025hypertension,aune2023blood,wang2023prevalence,gomez2017causes}. In hypertensive individuals, the relative risk of developing AF is significantly higher, with studies indicating a 50\% increased risk compared to normotensive patients \cite{aune2023blood}. This risk is further amplified by the degree of blood pressure elevation; even modest increases in systolic and diastolic blood pressure are associated with heightened AF risk. For example, a 20 mmHg increase in systolic blood pressure is linked to an 18\% increased risk of AF, while a 10 mmHg rise in diastolic pressure corresponds to a 7\% increase in risk \cite{aune2023blood}. In ICU settings, where patients are often acutely ill and under intensive monitoring, the presence of AF in hypertensive patients significantly complicates their clinical management, leading to poorer prognoses and extended ICU stays. Furthermore, the presence of AF in hypertensive ICU patients is associated with a substantially higher risk of all-cause mortality. Hypertensive patients with AF in the ICU exhibit a 2.55-fold increased risk of mortality compared to those without AF \cite{wang2023prevalence}. In broader hypertensive populations, AF has been shown to increase the risk of all-cause mortality by 1.5 to 2-fold \cite{gomez2017causes}. These findings underscore the urgent need for early identification and management of AF in hypertensive ICU patients to improve clinical outcomes and reduce the burden on healthcare resources. Given the multifactorial nature of mortality risk in these patients, early prediction of mortality is critical to guide timely and targeted interventions, optimize resource allocation, and enhance clinical decision-making. In particular, developing accurate and reliable prediction models for mortality risk in this high-risk group is crucial. Such models would not only help identify patients at the highest risk of adverse outcomes but also enable personalized treatment strategies, leading to better patient outcomes and more efficient use of healthcare resources.

While much of the research on hypertension and AF has focused on the development and progression of AF in hypertensive individuals, fewer studies have investigated its impact on clinical outcomes, particularly mortality, especially in critically ill ICU patients.
For example, Verdecchia et al. (2003) conducted a 16-year cohort study with 2,482 untreated hypertensive individuals to investigate the incidence and predictors of AF\cite{verdecchia2003atrial}. They found that age and left ventricular mass were the primary predictors of AF, with higher left ventricular mass increasing the risk by 1.20 times per standard deviation increase. Similarly, Huang et al. (2022) examined 412 patients to explore the relationship between ambulatory blood pressure (BP) patterns, particularly reverse dipping, and the development of AF\cite{huang2022prediction}. Their study revealed that reverse dippers were significantly more likely to have AF, and night-time BP patterns were stronger predictors of AF risk. Furthermore, Li et al. (2025) identified independent risk factors for AF within one year after discharge in hypertensive patients\cite{li2025clinical}. They found that male gender, lipoprotein(a), HbA1c, neutrophil-to-lymphocyte ratio, and triglyceride-glucose index were significant predictors, and developed a nomogram demonstrating a robust predictive accuracy with an Area Under the Receiver Operating Characteristic (AUROC) of 0.793.

Most studies to date have primarily concentrated on the factors leading to AF in hypertensive patients, but have not fully explored which influential factors, combined with the presence of AF, affect mortality risk in the ICU setting. This gap exists because traditional studies often focus on specific physiological or demographic factors that predict AF or its progression, rather than considering the complexity of mortality risk, which involves a wider range of dynamic and interrelated variables. Additionally, in ICU settings patients are typically critically ill, so the multifactorial nature of their risk factors make it challenging to predict outcomes using conventional statistical methods. Another reason is that the combination of AF and hypertension presents a particularly intricate risk profile that traditional models often fail to capture, as both conditions affect multiple organ systems and interact with other comorbidities in ways that are difficult to model with standard techniques.

In recent years, the rapid advancement of machine learning (ML) techniques has provided novel avenues for clinical prediction, owing to their ability to model complex, non-linear relationships within high-dimensional clinical data and uncover patterns that may be indiscernible through traditional statistical methods. Among the various machine learning models, CatBoost has demonstrated superior performance in clinical prediction tasks, primarily due to: (1) its native support for categorical features without the need for extensive preprocessing; (2) strong resistance to overfitting through innovative regularization techniques; and (3) efficient training and inference speeds, even on large and heterogeneous datasets. For example, Li et al. (2025) developed an interpretable machine learning model to predict postoperative stroke in elderly SICU patients using a large cohort of 19,085 patients\cite{li2025predicting}. After rigorous preprocessing and feature selection, their
CatBoost model achieved an AUROC of 0.8868, with prior cerebrovascular disease, serum creatinine, and systolic blood pressure identified as the top predictors. Si et al. (2025) developed an interpretable machine learning model using a large public database to predict in-hospital mortality after ICU cardiac arrest, and the CatBoost model achieved AUROC values of 0.904 (training) and 0.868 (test)\cite{si2025retrospective}. Through SHapley Additive exPlanations (SHAP) analysis, the most influential predictors identified were Glasgow Coma Scale (GCS), serum lactate, and minimum arterial blood pressure, highlighting the importance of neurological and metabolic parameters in risk stratification. Another example is by Li et al. (2023), they developed and validated a CatBoost-based machine learning model for predicting cardiac surgery-associated acute kidney injury (CSA-AKI) using multiple databases\cite{li2023development}. The model demonstrated strong discrimination, with AUROCs of 0.85, 0.67, and 0.77 on the internal, external, and MIMIC-IV validation datasets, respectively. SHAP analysis identified preoperative NT-proBNP, blood urea nitrogen, prothrombin time, serum creatinine, total bilirubin, and age as key risk factors for CSA-AKI, while higher platelets, blood pressure, albumin, and body weight were associated with lower risk. Collectively, these studies highlight the substantial potential of machine learning techniques, particularly CatBoost in advancing clinical prediction tasks. By delivering improved predictive accuracy and model interpretability, these approaches provide valuable decision-support tools for modern clinical practice, thereby facilitating more informed and timely decision-making in the complex environment of the ICU for this high risk population.

This study introduces several key innovations that distinguish our approach to early-phase mortality risk prediction in ICU patients with concurrent hypertension and AF:
\begin{itemize}
    \item \textbf{Clinical and Medical Significance}: This is the first study to focus specifically on short-term mortality risk in ICU patients with both hypertension and AF, utilizing routinely available clinical data to provide actionable predictions within 24 hours of ICU admission. By addressing this high-risk subgroup, our work directly supports timely triage and individualized interventions in critical care settings.
    
    \item \textbf{Pipeline Advantages}: This study developed a robust multi-domain feature engineering strategy that integrates chart events, laboratory biomarkers, therapeutic interventions, and comorbidity indices. After quality control, missing data imputation, and feature selection guided by expert clinical knowledge, we achieved a concise, physiologically relevant feature set that is applicable in real-world ICU scenarios.
    
    \item \textbf{Addressing Class Imbalance}: The challenge of class imbalance was systematically tackled through outcome-weighted training embedded within a cross-validated pipeline, ensuring sensitivity to minority outcomes and enhancing model robustness across different data splits.
    
    \item \textbf{Model Performance}: Our CatBoost model achieved superior performance, with an AUROC of 0.889 (95\% CI: 0.840–0.924), demonstrating its effectiveness in predicting 30-day mortality in ICU patients with coexisting hypertension and AF. This was further validated through benchmarking against five baseline models, confirming its optimal balance of accuracy and interpretability.
    
    \item \textbf{Interpretability Insights}: Using advanced explainability methods—including an ablation study, SHAP, Accumulated Local Effects (ALE), and DREAM analysis—this study identified critical predictors of ICU mortality, such as Richmond-RAS Scale, pO\textsubscript{2}, CefePIME, and Invasive Ventilation. These findings provide actionable insights into the clinical management of this complex patient population.
\end{itemize}

\section*{Materials and methods}
\subsection*{Data Source and Research Design}

This study utilized the Medical Information Mart for Intensive Care IV (MIMIC-IV, version 2.2) database, a publicly available and fully de-identified electronic health record (EHR) resource jointly curated by the Massachusetts Institute of Technology and the Beth Israel Deaconess Medical Center. MIMIC-IV comprises granular clinical data from more than 60,000 intensive care unit admissions recorded between 2008 and 2019\cite{johnson2023mimic}. The dataset spans a diverse array of information, including demographic details, physiologic measurements, laboratory results, medications, procedures, and patient outcomes. Its structured schema, high temporal fidelity, and comprehensive coverage across ICU domains make it particularly well-suited for developing and validating machine learning models in critical care research. All analyses were performed in accordance with the MIMIC-IV data usage license, and no additional institutional review board (IRB) approval was required, as all data were previously de-identified to ensure patient privacy.

This retrospective cohort study aimed to develop a transparent and clinically grounded machine learning framework to predict 30-day in-hospital mortality among ICU patients with coexisting hypertension and AF. The analysis followed a structured, multi-stage pipeline consisting of cohort construction, data preprocessing, feature selection, class imbalance correction, model development, performance evaluation, and interpretability assessment.

Patient selection criteria were based on structured ICD diagnostic codes, vital sign documentation, and demographic constraints. The study population was restricted to adult patients (age $\geq$ 18) experiencing their first ICU admission, with documented diagnoses of both AF and hypertension, while patients with malignancy were excluded to reduce heterogeneity from terminal comorbidities. For each patient, variables from five clinical domains—chartevents, labevents, procedureevents, medications, and admission/demographic data—were extracted from the first 24 hours of ICU stay to ensure early-stage prediction relevance.

An initial pool of over 300 candidate features was subjected to quality-based filtering and domain-informed review. Missing values were imputed using k-nearest neighbors (KNN), categorical variables were encoded using smoothed target encoding, and continuous variables were standardized using z-score normalization. A two-step feature selection strategy was applied: features with poor coverage or low variance were first removed, followed by a mutual information-based ranking to identify variables with the strongest association to 30-day mortality.

To address the class imbalance between survivors and non-survivors, we incorporated a class-weighted training strategy, preserving the outcome distribution in evaluation sets. Six machine learning algorithms—including CatBoost, LightGBM, XGBoost, logistic regression, Naïve Bayes, and shallow neural networks—were trained under a stratified 5-fold cross-validation scheme, with hyperparameters optimized via grid search. Model performance was primarily evaluated using AUROC and supported by metrics such as accuracy, F1-score, sensitivity, specificity, PPV, and NPV. Confidence intervals were calculated using 2,000 bootstrap replicates.

Finally, model interpretability was assessed using a combination of t-tests, ablation analysis, SHAP, ALE, and DREAM posterior sampling, allowing both global and individualized explanations. This design ensured that the developed model was not only statistically rigorous and discriminative, but also clinically interpretable and practically applicable in ICU decision support.

\begin{figure}[H]
\begin{adjustwidth}{-2.25in}{0in}
    \centering
    \includegraphics[width=0.85\linewidth]{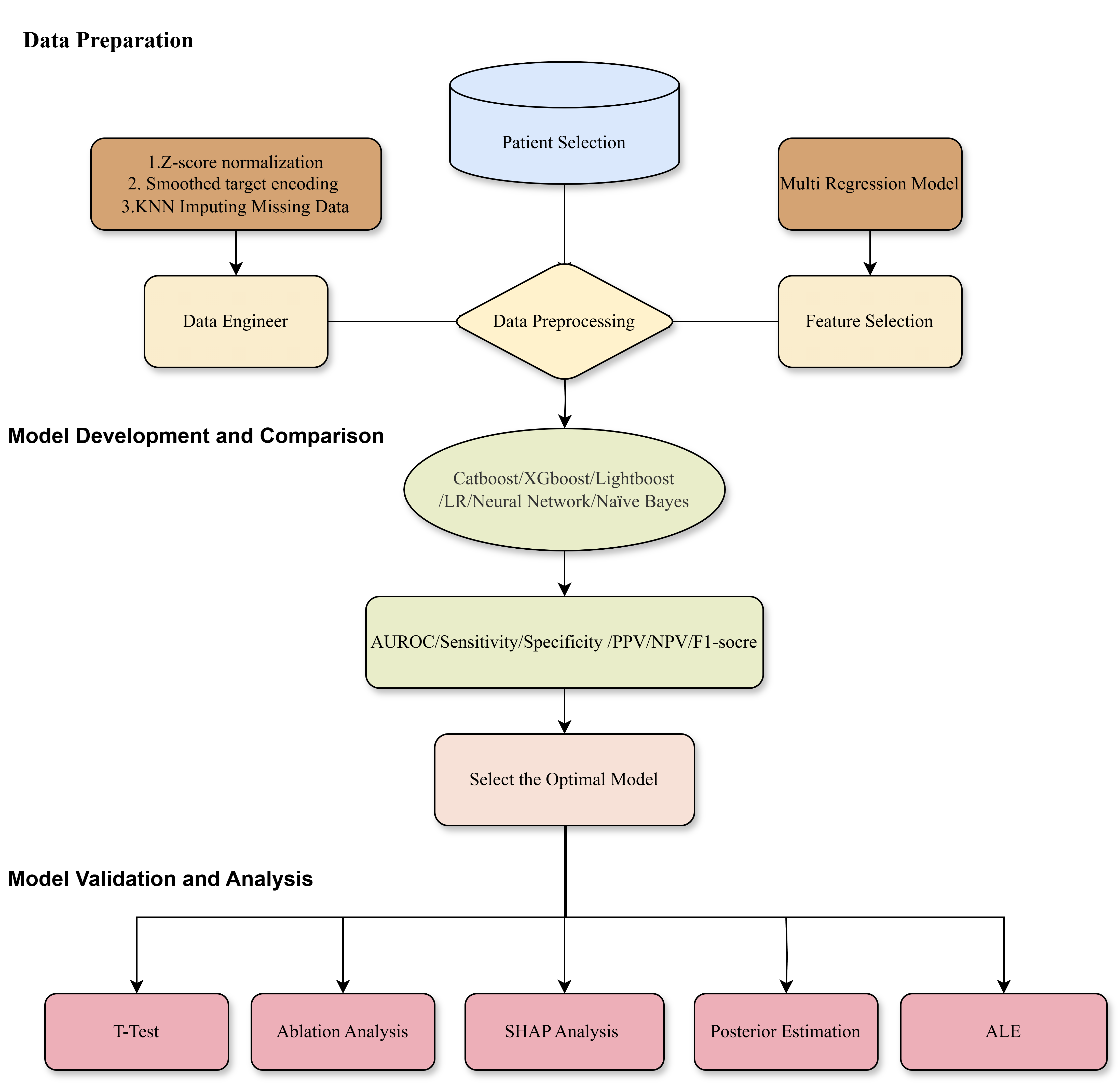}
    \caption{\textbf{Study design and modeling workflow overview.}}
    \label{fig:study_pipeline}
\end{adjustwidth}
\end{figure}

The complete study workflow is illustrated in Fig.~\ref{fig:study_pipeline}, ensuring transparency, methodological reproducibility, and relevance to real-world ICU applications.

\subsection*{Patient Selection}

This study aimed to develop a clinically interpretable model to predict 30-day in-hospital mortality among ICU patients aged 18 years and older with coexisting hypertension and AF. To construct a clinically meaningful and methodologically robust cohort, a series of structured inclusion and exclusion criteria were applied. Each step was designed to improve population homogeneity, minimize confounding, and enhance the model’s applicability to mortality risk stratification in critical care settings.

We began with 35{,}794 patients corresponding to their first ICU admission, ensuring that only the earliest ICU stay was analyzed per individual. This restriction reduces duplication bias and guarantees that the extracted features reflect the initial acute phase of critical illness.

\begin{figure}[H]
\centering
\includegraphics[width=0.5\linewidth]{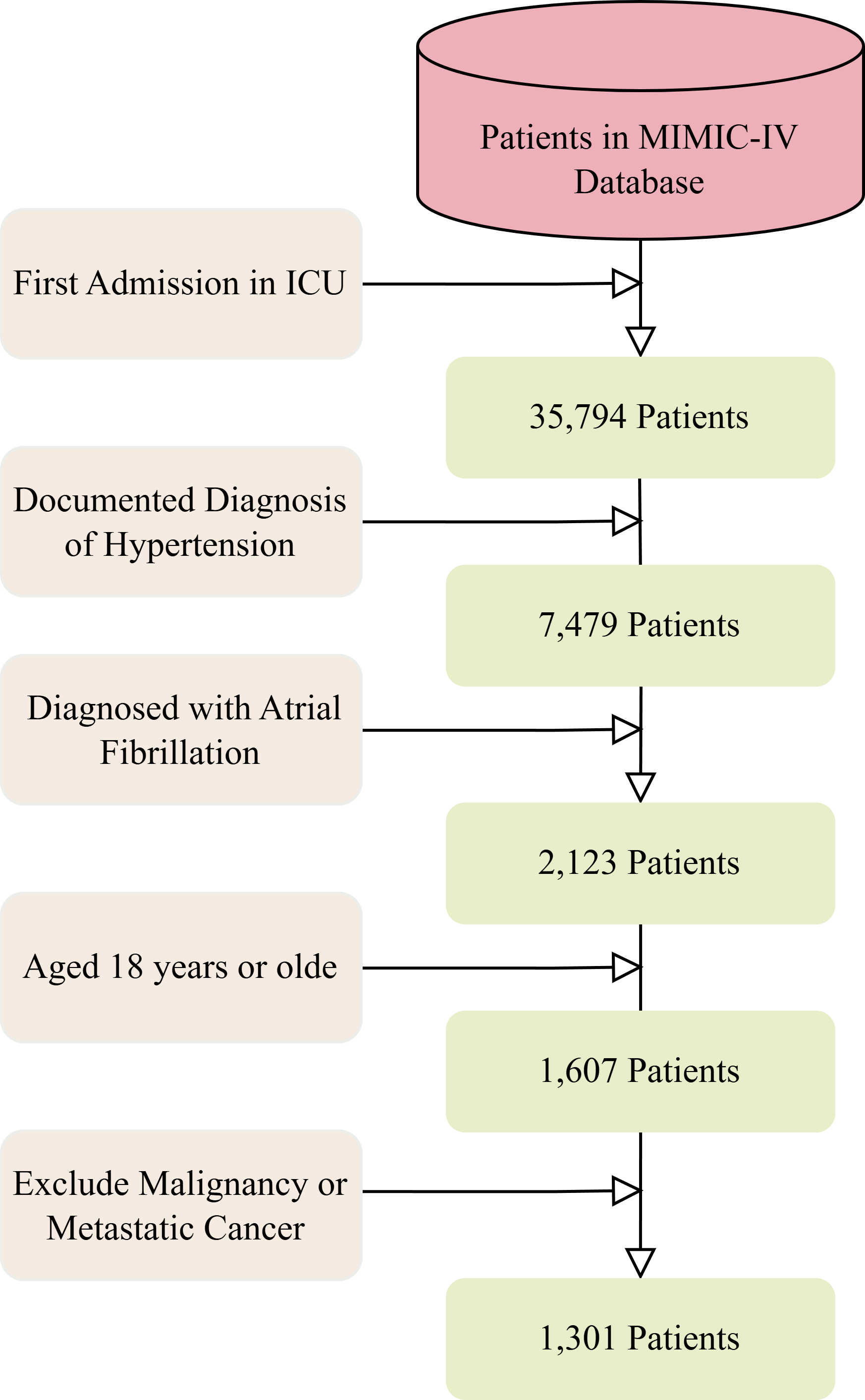} 
\caption{\textbf{Diagram depicting the patient selection procedure from MIMIC-IV.}}
\label{fig:patient_selection}
\end{figure}

As is shown in Fig.~\ref{fig:patient_selection}, from this population we identified 7{,}479 patients with a documented diagnosis of hypertension, as defined by standardized ICD-9 and ICD-10 codes. Hypertension is a prevalent and clinically impactful condition in the ICU population, contributing to increased cardiovascular risk and hemodynamic instability.

Among these, 2{,}123 patients were also diagnosed with atrial fibrillation—a common arrhythmia associated with elevated risks of stroke, systemic embolism, and ICU complications. The coexistence of hypertension and AF marks a high-risk subgroup deserving focused risk stratification.

To focus on the adult population, we applied an age filter to include only patients aged 18 years or older at the time of ICU admission ($n = 1{,}607$). This criterion excludes pediatric cases that follow different clinical trajectories and treatment protocols.

Finally, patients with active malignancies or metastatic cancer were excluded ($n = 1{,}301$), as their ICU prognosis is often governed by oncologic considerations and palliative care pathways. Excluding such patients ensures that the mortality predictions are specific to cardiovascular and critical care factors.

The final study cohort comprised 1{,}301 adult ICU patients with coexisting hypertension and atrial fibrillation, free of active malignancy, and undergoing their first ICU admission. For each patient, structured clinical data—including demographics, comorbidities, laboratory results, procedures, and interventions—were collected from the first 24 hours of ICU stay. The primary outcome was 30-day in-hospital mortality, defined by hospital discharge status and time to death.

\subsection*{Data preprocessing}
To ensure both predictive accuracy and clinical interpretability, we implemented a structured preprocessing framework that addresses common challenges in ICU datasets, including missingness, heterogeneous feature types, and class imbalance. Each method was selected to preserve medically meaningful signal while enhancing the stability and generalizability of downstream models.

Missing values were imputed using a KNN approach, which estimates missing entries by averaging values from the most similar patients in feature space\cite{pujianto2019k}. Compared to simple imputation techniques, KNN is better suited to high-dimensional clinical data with nonlinear dependencies. For instance, values missing from phosphorous or partial thromboplastin time (PTT)—both of which are not routinely measured in all ICU patients—can be reasonably estimated using related laboratory values such as anion gap and total bilirubin. The imputed value $\hat{x}_{i}^{(p)}$ for patient $p$ and feature $x_i$ is computed as:

\begin{equation}
\hat{x}_{i}^{(p)} = \frac{1}{k} \sum_{j \in \mathcal{N}_k(p)} x_i^{(j)}
\end{equation}

where $\mathcal{N}_k(p)$ denotes the set of $k$ nearest neighbors with non-missing values for $x_i$.

Categorical variables were encoded using smoothed target encoding\cite{nugroho2023smoothing}, which replaces each category with a regularized estimate of its mean outcome. This method is particularly appropriate for clinical features where categories have ordinal or interpretable associations with risk, such as Braden Nutrition or the presence of invasive ventilation. For example, patients with high Braden Nutrition scores tend to have better tissue perfusion and nutritional reserve, both of which are protective in critical illness. The encoded value for a category $c$ is calculated as:

\begin{equation}
\text{Encoded}(c) = \frac{n_c \cdot \bar{y}_c + \alpha \cdot \bar{y}}{n_c + \alpha}
\end{equation}

where $n_c$ is the number of samples with category $c$, $\bar{y}_c$ is the category-specific mortality rate, $\bar{y}$ is the global mean, and $\alpha$ is a smoothing factor.

All continuous variables were standardized using z-score normalization to account for scale differences across variables and stabilize model training\cite{si2025machine}. This was particularly important for features such as age, respiratory rate, and peak inspiratory pressure, which span different value ranges but are clinically comparable in their predictive importance. Standardization was performed as:

\begin{equation}
x_i^* = \frac{x_i - \mu_i}{\sigma_i}
\end{equation}

where $\mu_i$ and $\sigma_i$ are the mean and standard deviation of feature $x_i$ in the training set.

To address the imbalance in 30-day mortality outcomes, class weighting was applied during model training. This approach increases the penalty for misclassifying deaths, which are underrepresented in the dataset\cite{Chen2025.05.11.25327405}, and thus improves the model’s sensitivity to high-risk patients. The class weight $w_y$ for outcome class $y$ is defined as:

\begin{equation}
w_y = \frac{1}{f_y}
\end{equation}

where $f_y$ is the relative frequency of class $y$ in the training set. For example, higher weights were assigned to non-survivors to ensure that rare but clinically critical patterns—such as elevated charlson comorbidity index or antibiotic administration (e.g., cefepime)—were not overshadowed by more prevalent survival features.

All preprocessing steps, including imputation, encoding, and normalization, were conducted exclusively within each training fold of a stratified five-fold cross-validation pipeline. This prevents information leakage and ensures that model performance reflects generalizable patterns rather than artifacts of the data split. The resulting feature matrix reflects a clinically coherent, statistically consistent representation of patients with hypertension and atrial fibrillation in the ICU.

\subsection*{Feature selection}

To construct a clinically relevant and statistically efficient set of predictors, a structured feature selection strategy was employed, grounded in data quality, physiological relevance, and modeling requirements, and validated through expert endorsement, ensuring alignment with clinical expertise. Candidate features were categorized into five clinically intuitive domains: chart events, laboratory events, procedures, medications, and demographics/comorbidities. This categorization was guided by clinical input from senior intensivists, ensuring alignment with real-world ICU practice and the pathophysiological nuances of atrial fibrillation and hypertension. The final scheme reflects the multifactorial risk architecture of ICU patients with AF and hypertension, in whom mortality risk arises from organ dysfunction, therapeutic interventions, and chronic health status.

\textbf{Chart events} included high-frequency bedside observations and nursing assessments. Variables such as BUN, Braden Nutrition, Peak Inspiratory Pressure, and Anion gap were selected due to their ability to capture renal perfusion, nutritional risk, respiratory mechanics, and acid-base imbalance—critical elements in AF patients with labile hemodynamics and multiorgan stress\cite{li2024association}.

\textbf{Laboratory events} consisted of systemic biomarkers such as pO\textsubscript{2}, which offers insight into oxygenation efficiency and respiratory compromise\cite{smith2018atrial}. These values often reflect the severity of pulmonary dysfunction, which frequently coexists with cardiovascular instability in this population.

\textbf{Procedure events} recorded the use of invasive support, with Invasive Ventilation serving as a proxy for respiratory failure severity and treatment intensity\cite{lyons2017design}. The inclusion of this variable supports model sensitivity to advanced ICU interventions that often define prognosis.

\textbf{Drug administration} data captured therapeutic intensity, including the use of broad-spectrum antibiotics such as CefePIME\cite{chanderraj2024mortality}. Antibiotic exposure can signal suspected or confirmed infection, a known exacerbator of AF burden and mortality risk in hypertensive patients.

\textbf{Demographics and comorbidities} included age and the Charlson comorbidity index, which together quantify baseline vulnerability\cite{jung2024multimorbidity}. Age is a well-established determinant of ICU outcomes, and the Charlson index aggregates chronic disease burden with strong prognostic value.

To refine the initial candidate list, we implemented a two-step selection strategy. The process began with an initial pool of approximately 300 candidate features derived from structured EHR fields across multiple MIMIC-IV tables. In the first stage, variables with more than 20\% missingness or documented in fewer than 100 unique patient records were excluded. This threshold was set to ensure the reliability of feature estimates and to mitigate the risk of model instability resulting from excessive data sparsity. During this stage, features such as \textit{GCS}, \textit{Chloride}, and \textit{Glucose} were removed due to poor documentation frequency or highly inconsistent recording patterns in the ICU setting.

In the second stage, we employed mutual information (MI) ranking to quantify each feature's dependency on the binary outcome of 30-day in-hospital mortality. MI is a non-parametric method that captures both linear and nonlinear associations, making it particularly well-suited for heterogeneous ICU data:

\begin{equation}
MI(x; y) = \sum_{x \in X} \sum_{y \in Y} p(x, y) \cdot \log \frac{p(x, y)}{p(x)p(y)}
\end{equation}

Features with near-zero MI scores—such as \textit{Oxygen Device Type} and \textit{Hospital Admission Source}—were excluded due to negligible contribution to outcome variability. These variables were either clinically redundant, inconsistently defined across ICU units, or failed to differentiate survivors from non-survivors in a statistically meaningful way.

As is shown in Table~\ref{tab:final_features_af_hyp}, the final feature set comprised 17 variables, selected for their clinical interpretability, data availability, and relevance to AF and hypertensive pathophysiology. This curated set spans major physiologic axes and reflects both acute illness manifestations and chronic vulnerability factors. The final selection was confirmed in consultation with experienced critical care specialists to ensure that all retained variables possessed clear clinical significance and reflected actionable information in ICU settings.

\begin{table}[htbp]
\centering
\caption{\textbf{The 17 selected features for predicting 30-day mortality in ICU patients with hypertension and AF.}}
\label{tab:final_features_af_hyp}
\small
\renewcommand{\arraystretch}{1.2}
\begin{tabularx}{\textwidth}{>{\raggedright\arraybackslash}p{4.5cm}|>{\raggedright\arraybackslash}X}
\hline
\rowcolor[HTML]{f7e1d7}
\textbf{Category} & \textbf{Selected Features} \\
\hline
Chart events & BUN, Richmond-RAS Scale, PTT, Phosphorous, Total Bilirubin, Anion gap, Differential-Lymphs, Braden Nutrition, Braden Moisture, Respiratory Rate (Set), Activity/Mobility (JH-HLM), Peak Inspiratory Pressure \\
\hline
Laboratory events & pO\textsubscript{2} \\
\hline
Procedure events & Invasive Ventilation \\
\hline
Drug administration & CefePIME \\
\hline
Admission/Demographics & Age, Charlson comorbidity index \\
\hline
\end{tabularx}
\end{table}

\subsection*{Model Development}

\begin{figure}[H]
    \centering
    \includegraphics[width=1.0\linewidth]{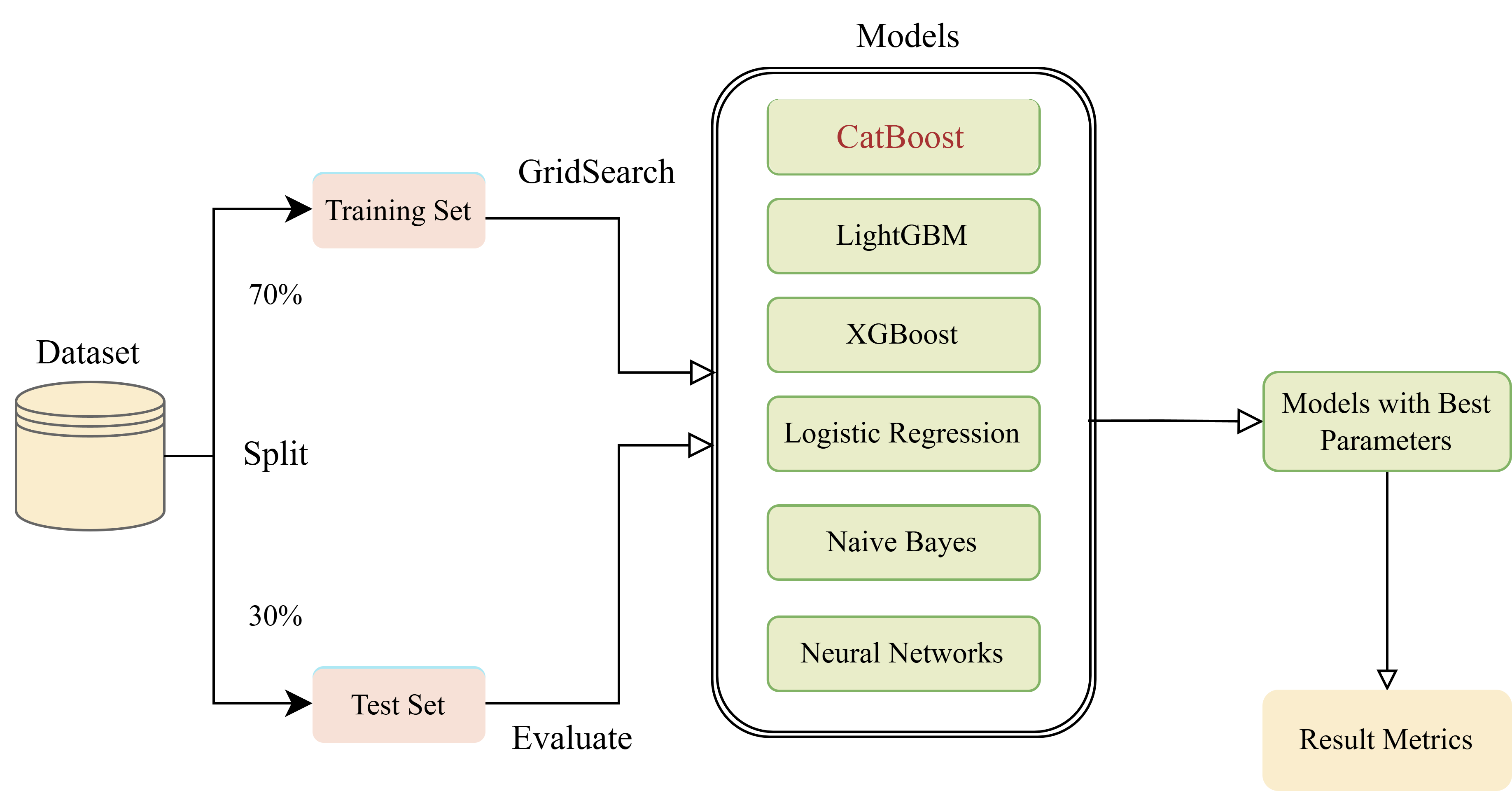}
    \caption{\textbf{Machine learning models used to predict 30-day in-hospital mortality in ICU patients with hypertension and AF.}}

    \label{fig:model}
\end{figure}

 We established a supervised machine learning framework encompassing diverse model architectures. Following feature construction and cohort stratification (70\% training, 30\% testing), six classification models were trained and fine-tuned via stratified five-fold cross-validation using grid search. The overall modeling strategy is depicted in Fig.~\ref{fig:model}.

\newpage
Among the models, \textbf{gradient boosting techniques}—including CatBoost, LightGBM, and XGBoost—were selected for their ability to capture intricate nonlinear dependencies and interaction effects within high-dimensional clinical data\cite{liu2021prediction}. These ensemble-based learners construct additive tree sequences, progressively minimizing prediction error. CatBoost was advantageous due to its support for categorical features and mitigation of overfitting through ordered boosting mechanisms. In contrast, LightGBM and XGBoost adopt efficient tree-growing strategies, with hyperparameters such as learning rate, tree depth, subsample ratio, and regularization terms (\texttt{reg\_alpha}, \texttt{reg\_lambda}) optimized for stability and accuracy.

\textbf{Logistic regression} was employed as a transparent and interpretable baseline. Both L1 and L2 penalty schemes were tested to balance bias-variance tradeoffs and reduce coefficient inflation due to multicollinearity. The regularization strength parameter \( C \) was determined via nested cross-validation. While inherently linear, this model offered valuable clinical interpretability through direct coefficient analysis.

\textbf{Naive Bayes} provided a fast, probabilistic benchmark. Assuming conditional independence, this classifier modeled class-conditional likelihoods using a Gaussian distribution for continuous variables. Although not flexible enough for complex ICU settings, it served as a computationally efficient point of comparison.

A \textbf{neural network} was also developed, comprising one hidden layer with ReLU activation and a sigmoid output layer. The network was trained using the Adam optimizer and binary cross-entropy loss, with hyperparameters (e.g., learning rate, batch size, dropout rate) adjusted to ensure convergence. While neural networks offer expressive modeling capacity, their reduced interpretability and potential overfitting in limited datasets must be considered in clinical applications.

Each model was trained with the goal of maximizing predictive performance while minimizing overfitting. Appropriate regularization strategies—such as L1/L2 penalties, dropout, early stopping, and tree-based constraints—were applied according to each algorithm’s structure to enhance generalizability.

To comprehensively assess model behavior, we adopted a multidimensional evaluation framework incorporating several clinically meaningful metrics. While AUROC was used as the primary measure of overall discriminative ability due to its threshold-independent nature, additional metrics were essential for understanding clinical applicability. Accuracy, though intuitive, may overestimate performance in imbalanced datasets such as ICU mortality, where survivors are the majority class. Therefore, F1-score was included to balance precision and recall, ensuring fair treatment of false positives and false negatives—both of which carry serious consequences in critical care. Sensitivity (true positive rate) was emphasized to evaluate the model’s ability to detect high-risk patients, thereby minimizing missed critical cases. Conversely, specificity (true negative rate) ensured that low-risk individuals were not over-identified, reducing unnecessary clinical interventions. Positive predictive value (PPV) quantified the reliability of high-risk predictions, informing clinicians about the likelihood of actual adverse outcomes when alerts are triggered, while negative predictive value (NPV) provided confidence when ruling out risk, which is particularly important for discharge decisions or de-escalation of care. Confidence intervals for AUROC were estimated using 2{,}000 bootstrap replicates to quantify statistical robustness. This ensemble of metrics enabled a balanced and nuanced evaluation, offering both statistical rigor and clinical relevance for model selection in the context of mortality prediction among ICU patients with atrial fibrillation and hypertension.

\subsection*{Statistical Analyses}

To ensure methodological integrity, clinical interpretability, and operational trustworthiness of the predictive framework, we implemented a multi-level statistical validation and explanation pipeline. Each method was chosen to address a specific modeling challenge—ranging from cohort comparability and feature evaluation to model transparency and personalized risk estimation—ensuring that the predictive system was not only statistically sound, but also actionable in critical care contexts.

We began by validating the baseline comparability between training and test cohorts. Two-sided independent sample t-tests were conducted on key clinical variables (e.g., age, anion gap, BUN), with Welch’s correction applied when variances were unequal\cite{sun2025optimizing}. The t-statistic was computed as:

\begin{equation}
t = \frac{\bar{x}_1 - \bar{x}_2}{\sqrt{\frac{s_1^2}{n_1} + \frac{s_2^2}{n_2}}},
\end{equation}

This step ensured that stratified random sampling did not induce distributional shifts, preserving generalizability across ICU subgroups and reducing the risk of biased performance estimates.

To isolate and interpret each feature’s marginal predictive utility, we conducted an ablation analysis by iteratively removing one feature at a time from the final model and re-evaluating the AUROC.
This approach enabled direct measurement of the performance impact of each variable, supporting both model refinement and clinical prioritization. For instance, features related to organ dysfunction (e.g.,pO\textsubscript{2}, CefePIME) demonstrated high marginal importance consistent with their pathophysiological roles in AF and hypertensive ICU patients.

To understand how feature changes influence model predictions beyond linear assumptions, we applied Accumulated Local Effects \cite{bischof2008should}. Unlike partial dependence plots, ALE is robust to multicollinearity and respects the empirical distribution of covariates. For a feature \(x_j\), ALE is defined as:

\begin{equation}
ALE_j(z) = \int_{z_0}^{z} \mathbb{E}_{X_{\setminus j}} \left[ \frac{\partial f(X)}{\partial x_j} \Big| x_j = s \right] ds,
\end{equation}

This allowed us to characterize non-monotonic or threshold-like effects common in ICU physiology, such as U-shaped risk relationships observed in respiratory metrics or biochemical markers. ALE plots provided clinically meaningful patterns that could inform decision thresholds or care escalation points.

To complement global insights, we further employed SHAP to decompose each individual prediction into feature-level contributions\cite{ashrafi2024optimizing}. The SHAP value $\phi_i$ is defined as:

\begin{equation}
\phi_i = \sum_{S \subseteq F \setminus \{x_i\}} \frac{|S|! (|F|-|S|-1)!}{|F|!} \left[f(S \cup \{x_i\}) - f(S)\right],
\end{equation}

SHAP values offer local, additive attributions grounded in cooperative game theory, allowing for transparent interpretation of individual predictions—especially critical for clinical users seeking explanation for high-risk classifications.

Finally, to quantify individual-level uncertainty in risk prediction, we incorporated a posterior estimation framework using the DiffeRential Evolution Adaptive Metropolis algorithm\cite{singh1986organizational}. By integrating over the posterior distribution of model parameters, we obtained calibrated confidence intervals for each predicted mortality probability:

\begin{equation}
p(y \mid X) = \int p(y \mid \theta, X) \cdot p(\theta \mid X) \, d\theta,
\end{equation}

This capability is vital in ICU environments where borderline risk cases demand not only a score but also an estimate of confidence or ambiguity—enabling clinicians to appropriately triage and monitor patients.

Together, this ensemble of statistical techniques ensured that our model was not only predictive and generalizable, but also interpretable, robust to confounding, and aligned with clinical reasoning. The integration of feature-level and patient-level explanations alongside uncertainty quantification supports both operational deployment and ethical transparency in real-world ICU applications.

\section*{Results}
\subsection*{Cohort Balance and Clinical Differentiation}

To evaluate the robustness and clinical interpretability of the 30-day mortality prediction model developed for ICU patients with comorbid hypertension and AF, we conducted a two-part statistical validation. First, we examined the consistency of clinical features between the training ($n=911$) and test sets ($n=390$), followed by a comparison between survival outcomes to ensure that the model captures clinically meaningful mortality stratifiers.

As shown in Table~\ref{tab:cohort comparison results}, the distributions of key variables—including age (69.74 vs 69.52 years, $p=0.687$), PTT (36.24 vs 36.44 s, $p=0.827$), RR\textsubscript{set} (18.19 vs 18.06 bpm, $p=0.531$), and peak inspiratory pressure (19.38 vs 19.49 cmH\textsubscript{2}O, $p=0.630$)—were not significantly different between cohorts. Categorical features such as use of invasive ventilation and exposure to CefePIME also maintained similar prevalence (both $p>0.5$). These findings support the structural integrity of the modeling pipeline and suggest that the generalization from training to unseen cases is unlikely to be confounded by data shift.

Conversely, comparison between survivors and non-survivors (Table~\ref{tab:cohort comparison results 1}) revealed significant physiological and clinical distinctions. Markers of organ dysfunction, such as elevated BUN (40.40 vs 20.71 mg/dL, $p<0.001$), higher PTT (47.19 vs 34.87 s, $p<0.001$), and increased bilirubin (2.78 vs 1.21 mg/dL, $p=0.008$), were all enriched in non-survivors. These patterns indicate worsening renal and hepatic function among those who died. Importantly, indicators of respiratory burden, such as higher RR\textsubscript{set} (20.49 vs 17.90 bpm, $p<0.001$) and peak inspiratory pressure (21.70 vs 19.09 cmH\textsubscript{2}O, $p<0.001$), aligned with more aggressive ventilation support in the fatal cohort.

Functional and frailty-associated variables also demonstrated predictive value. The Richmond-RAS scale was markedly lower in non-survivors ($-2.50$ vs $-0.90$, $p<0.001$), reflecting deeper sedation or reduced consciousness. Nutritional and mobility scores (Braden-Nutr: 2.01 vs 2.44, $p<0.001$; JH-HLM: 2.00 vs 2.39, $p<0.001$) were similarly worse in those who died, capturing the underlying physical vulnerability often exacerbated in older HTN-AF populations.

Together, these results confirm the model's capacity to generalize across cohorts and reflect clinically interpretable drivers of ICU mortality. The tight alignment between statistical outputs and medical rationale reinforces the potential application of this model in triage decision-making for complex patients with HTN and AF.

\begin{table}[H]
\noindent
\caption{\textbf{T-test analysis between the training and test datasets.}}
\label{tab:cohort comparison results}
\small
\renewcommand{\arraystretch}{1.2}
\rowcolors{2}{white}{white}
\begin{tabular}{lllll}
\toprule
Feature & Unit & Training Set& Test Set& P-value \\
\midrule
Richmond-RAS Scale & – & -1.08 (1.30) & -1.01 (1.30) & 0.352 \\
BUN & mg/dL & 22.90 (17.85) & 20.03 (11.82) & \textless{} 0.001 \\
PTT & sec & 36.24 (15.45) & 36.44 (14.60) & 0.827 \\
pO\textsubscript{2} & mmHg & 153.23 (68.96) & 160.07 (72.61) & 0.114 \\
Braden Nutrition & score & 2.39 (0.45) & 2.42 (0.45) & 0.271 \\
Total Bilirubin & mg/dL & 1.38 (2.24) & 1.28 (2.02) & 0.411 \\
Activity / Mobility (JH-HLM) & score & 2.34 (0.51) & 2.40 (0.54) & 0.086 \\
Phosphorous & mg/dL & 3.48 (1.01) & 3.51 (1.02) & 0.318 \\
Anion gap & mEq/L & 12.97 (3.35) & 12.52 (3.21) & 0.023 \\
Respiratory Rate (Set) & breaths/min & 18.19 (3.36) & 18.06 (3.11) & 0.531 \\
Peak Inspiratory Pressure & cmH\textsubscript{2}O & 19.38 (4.13) & 19.49 (3.74) & 0.630 \\
Braden Moisture & score & 3.55 (0.40) & 3.58 (0.38) & 0.202 \\
Age & years & 69.74 (9.31) & 69.52 (8.88) & 0.687 \\
Differential-Lymphs & \% & 13.82 (6.93) & 14.28 (6.85) & 0.267 \\
Charlson Comorbidity Index & score & 4.28 (1.67) & 4.28 (1.77) & 0.983 \\
CefePIME & presence & 0.28 (0.45) & 0.26 (0.44) & 0.531 \\
Invasive Ventilation & presence & 0.48 (0.50) & 0.46 (0.50) & 0.669 \\
\bottomrule
\end{tabular}
\begin{flushleft}
\textbf{Table notes}: Descriptive statistics for clinical variables in the training and test cohorts. Values are presented as mean (SD), and statistical significance was set at a threshold of 0.05.
\end{flushleft}
\end{table}

\begin{table}[H]
\noindent
\caption{\textbf{T-test analysis between the survival and non-survival groups.}}
\label{tab:cohort comparison results 1}
\small
\renewcommand{\arraystretch}{1.2}
\rowcolors{2}{white}{white}
\begin{tabular}{lllll}
\toprule
Feature & Unit & Survivors & Non-Survivors & P-value \\
\midrule
Richmond-RAS Scale & – & -0.90 (1.13) & -2.50 (1.68) & \textless{} 0.001 \\
BUN & mg/dL & 20.71 (13.50) & 40.40 (32.81) & \textless{} 0.001 \\
PTT & sec & 34.87 (13.43) & 47.19 (24.01) & \textless{} 0.001 \\
pO\textsubscript{2} & mmHg & 159.96 (68.61) & 99.30 (43.49) & \textless{} 0.001 \\
Braden Nutrition & score & 2.44 (0.43) & 2.01 (0.40) & \textless{} 0.001 \\
Total Bilirubin & mg/dL & 1.21 (1.09) & 2.78 (5.82) & 0.008 \\
Activity / Mobility (JH-HLM) & score & 2.39 (0.51) & 2.00 (0.30) & \textless{} 0.001 \\
Phosphorous & mg/dL & 3.39 (0.88) & 4.12 (1.57) & \textless{} 0.001 \\
Anion gap & mEq/L & 12.67 (3.07) & 15.33 (4.42) & \textless{} 0.001 \\
Respiratory Rate (Set) & breaths/min & 17.90 (2.98) & 20.49 (4.95) & \textless{} 0.001 \\
Peak Inspiratory Pressure & cmH\textsubscript{2}O & 19.09 (3.74) & 21.70 (5.98) & \textless{} 0.001 \\
Braden Moisture & score & 3.59 (0.38) & 3.25 (0.42) & \textless{} 0.001 \\
Age & years & 69.64 (9.21) & 70.54 (10.11) & 0.396 \\
Differential-Lymphs & \% & 14.34 (6.83) & 9.66 (6.33) & \textless{} 0.001 \\
Charlson Comorbidity Index & score & 4.24 (1.70) & 4.59 (1.44) & 0.024 \\
CefePIME & presence & 0.24 (0.42) & 0.63 (0.48) & \textless{} 0.001 \\
Invasive Ventilation & presence & 0.47 (0.50) & 0.55 (0.50) & 0.096 \\
\bottomrule
\end{tabular}
\begin{flushleft}
\textbf{Table notes}: This table compares patients who survived versus those who died within 30 days. It presents the differences in mean values for key clinical variables along with p-values, with statistical significance set at a threshold of 0.05.
\end{flushleft}
\end{table}

\subsection*{Ablation Study and Model Robustness in Hypertension-AF Cohort}

\begin{figure}[H]
\begin{adjustwidth}{-2.25in}{0in}
    \centering
    \includegraphics[width=1.05\linewidth]{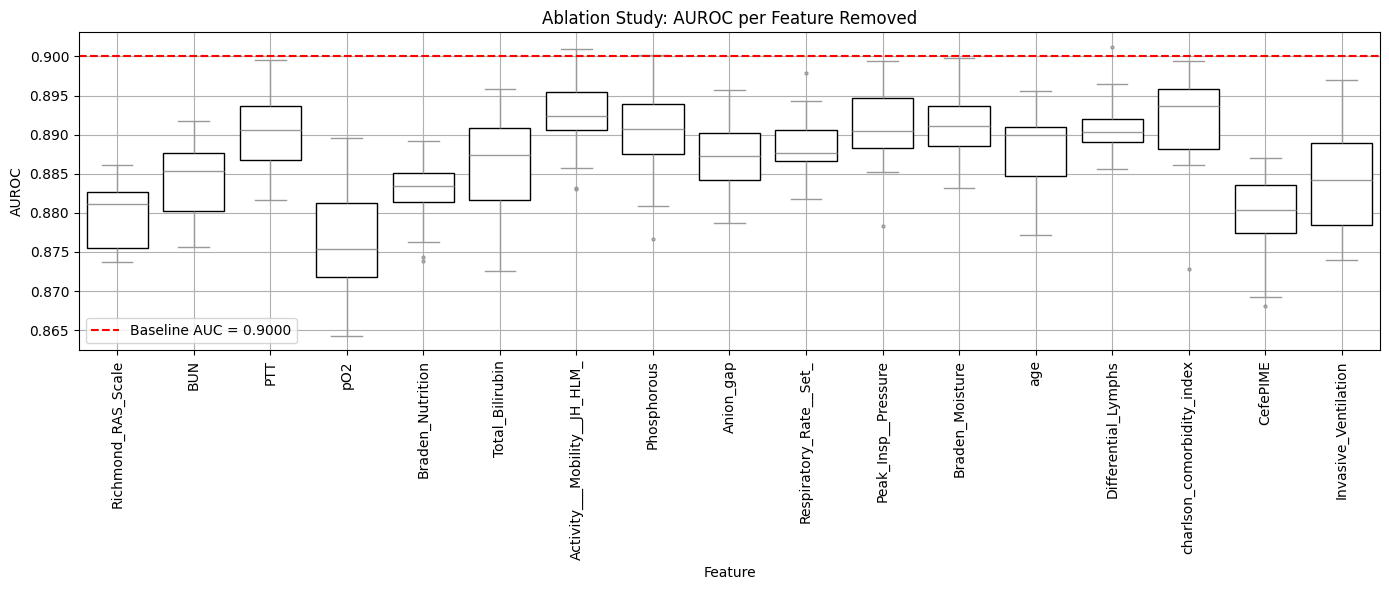}
    \caption{\textbf{Effect of Feature Exclusion on CatBoost Model Performance.}}
    \label{fig:ablation analysis}
\end{adjustwidth}
\end{figure}

To evaluate the reliability of the model in predicting mortality at 30 days among patients admitted to the ICU for atrial fibrillation (AF) in the context of preexisting hypertension, we performed an ablation analysis using logistic regression. As illustrated in Fig.~\ref{fig:ablation analysis}, each clinical characteristic was removed sequentially and the model was re-trained and tested in bootstrap replicates. The resulting distribution of AUROC values highlights the contribution of each variable to overall discrimination ability.

Across the ablation spectrum, most features exhibited limited individual perturbation effects, with AUROC values remaining tightly clustered around the baseline (0.9000). This pattern indicates that the model does not over-rely on any single input, and instead benefits from a distributed representation across diverse physiological domains. Notably, the AUROC variance remained low for nearly all ablations, further supporting the stability of the logistic formulation under feature omission.

Although removal of certain features such as \textit{pO\textsubscript{2}}, \textit{CefePIME}, or \textit{Invasive Ventilation} led to mild fluctuations in predictive performance, the absence of a specific feature did not result in a catastrophic drop, reinforcing the redundancy and generalizability of the model. Such statistical resilience, even in a clinically complex population such as admissions to the hypertensive AF ICU, underscores the potential for reliable deployment in real-world settings.

In general, the ablation results validate both the structural robustness and the interpretability of the model. The findings support the notion that logistic regression, when carefully specified and evaluated, remains a competitive and transparent choice for risk prediction in critical care applications.

\subsection*{Model Performance Evaluation and Clinical Interpretability}

\begin{figure}[H]
\centering
\includegraphics[width=0.88\linewidth]{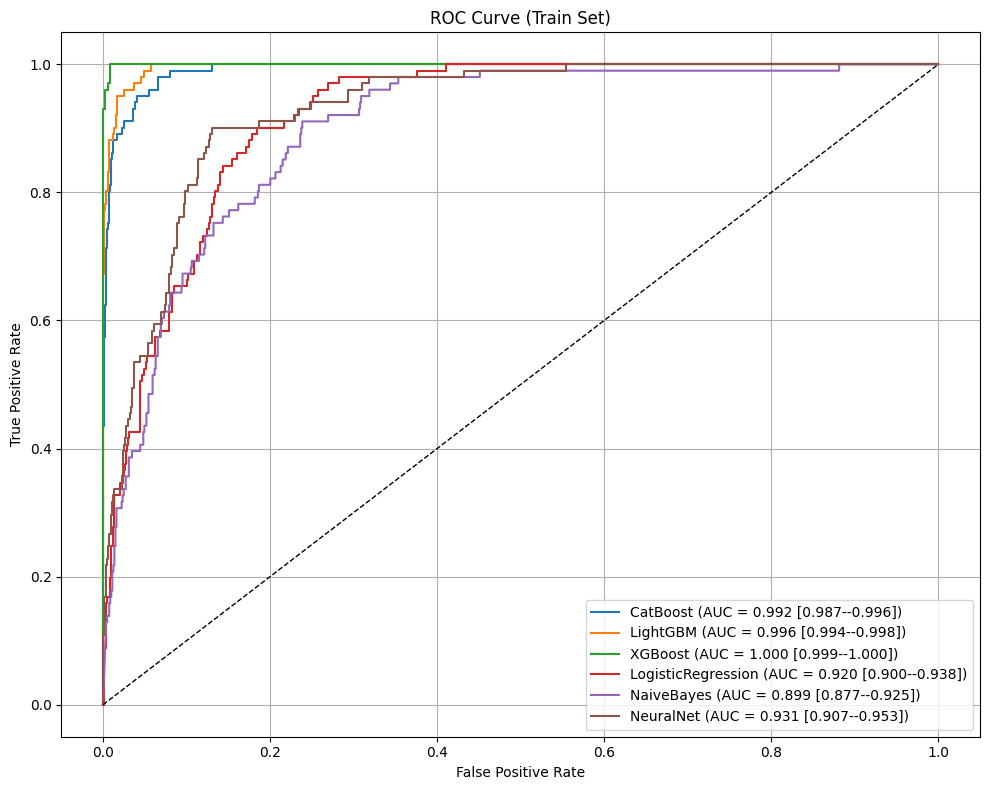}
\caption{\textbf{AUROC Curves for Model Performance in the Training Set.}}
\label{fig:roc_train}
\end{figure}

\begin{figure}[H]
\centering
\includegraphics[width=0.88\linewidth]{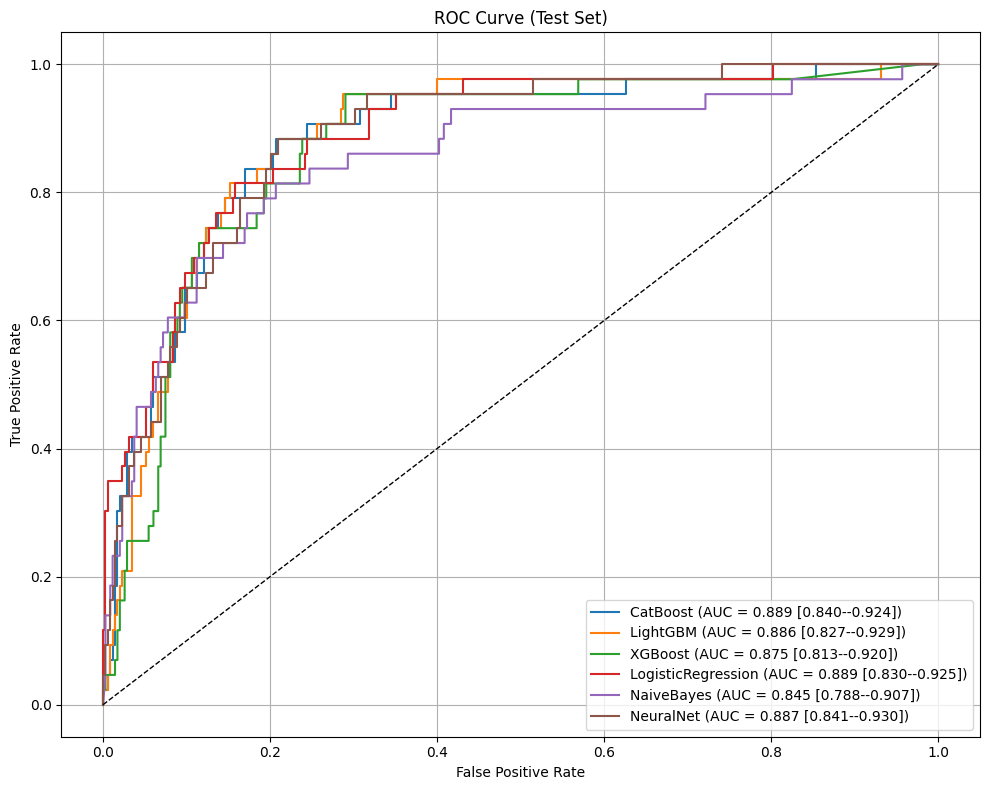}
\caption{\textbf{AUROC Curves for Model Performance in the Test Set.}}
\label{fig:roc_test}
\end{figure}

To evaluate the discriminative power and generalizability of machine learning models for predicting 30-day mortality among ICU patients with coexisting hypertension and atrial fibrillation, we conducted a comparative analysis of six representative algorithms. Table~\ref{tab: Results of the Training Set} and Table~\ref{tab: Results of the Test Set} report key performance metrics, and Fig.~\ref{fig:roc_train} and Fig.\ref{fig:roc_test} illustrate the ROC curves on the training and test sets, respectively.

On the training set, all models achieved strong classification performance, with XGBoost reaching an AUROC of 1.000, followed closely by LightGBM (0.996) and CatBoost (0.992). While this may suggest potential overfitting, model behavior on the test set confirms robustness. CatBoost and logistic regression both maintained an AUROC of 0.889(95\%CI:0.840-0.924) on unseen data, with LightGBM performing similarly (0.886). Notably, all models preserved a high sensitivity around 0.814–0.837 after threshold adjustment, ensuring that patients at high risk of death were consistently identified. At the same time, specificity values remained above 0.80, preserving reliability in ruling out low-risk cases.

Despite neural networks and Naive Bayes presenting modest discrimination in the test set (AUROC of 0.887 and 0.845, respectively), they still maintained high negative predictive values ($NPV > 0.97$), reinforcing clinical safety in low-risk stratification. Logistic regression served as a baseline with competitive performance, validating the stability of linear predictors in this setting.

The consistent ranking of clinically relevant features across all models—such as blood urea nitrogen, respiratory parameters, and functional mobility—underscores biological plausibility and suggests that predictive insights are rooted in established pathophysiological processes. The relatively low performance drop from training to test phases further supports the generalizability of these models in heterogeneous ICU populations.

In summary, CatBoost offered an optimal balance of accuracy, interpretability, and generalization in predicting 30-day mortality in hypertensive ICU patients with atrial fibrillation. Additionally, the strong and consistent performance of other machine learning models further enhanced the overall robustness and interpretability of the predictive framework.

\begin{table}[H]
\small
\renewcommand{\arraystretch}{1.2}
\begin{adjustwidth}{-2.25in}{0in}
\centering
\caption{\textbf{Performance Comparison of Different Models in the Training Set.}}
\begin{tabular}{l|l|l|l|l|l|l|l}
\hline
\rowcolor[HTML]{f7e1d7}
\textbf{Model} & \textbf{AUROC (95\% CI)} & \textbf{Accuracy} & \textbf{F1-score} & \textbf{Sensitivity} & \textbf{Specificity} & \textbf{PPV} & \textbf{NPV} \\ \hline
\rowcolor[HTML]{a8dadc}
CatBoost & 0.992 (0.987--0.996) & 0.960 & 0.839 & 0.931 & 0.964 & 0.764 & 0.991 \\
LightGBM & 0.996 (0.994--0.998) & 0.976 & 0.897 & 0.950 & 0.979 & 0.850 & 0.994 \\
XGBoost & 1.000 (0.999--1.000) & 0.991 & 0.962 & 0.990 & 0.991 & 0.935 & 0.999 \\
LogisticRegression & 0.920 (0.900--0.938) & 0.851 & 0.556 & 0.842 & 0.852 & 0.415 & 0.977 \\
NaiveBayes & 0.899 (0.877--0.925) & 0.860 & 0.538 & 0.733 & 0.876 & 0.425 & 0.963 \\
Neural Networks & 0.931 (0.907--0.953) & 0.824 & 0.535 & 0.911 & 0.813 & 0.379 & 0.987 \\ \hline
\end{tabular}
\label{tab: Results of the Training Set}
\end{adjustwidth}
\end{table}

\begin{table}[H]
\small
\renewcommand{\arraystretch}{1.2}
\begin{adjustwidth}{-2.25in}{0in}
\centering
\caption{\textbf{Performance Comparison of Different Models in the Test Set.}}
\begin{tabular}{l|l|l|l|l|l|l|l}
\hline
\rowcolor[HTML]{f7e1d7}
\textbf{Model} & \textbf{AUROC (95\% CI)} & \textbf{Accuracy} & \textbf{F1-score} & \textbf{Sensitivity} & \textbf{Specificity} & \textbf{PPV} & \textbf{NPV} \\ \hline
\rowcolor[HTML]{a8dadc}
CatBoost & 0.889 (0.840--0.924) & 0.831 & 0.522 & 0.837 & 0.830 & 0.379 & 0.976 \\
LightGBM & 0.886 (0.827--0.929) & 0.844 & 0.534 & 0.814 & 0.848 & 0.398 & 0.974 \\
XGBoost & 0.875 (0.813--0.920) & 0.806 & 0.479 & 0.814 & 0.805 & 0.340 & 0.972 \\
LogisticRegression & 0.889 (0.830--0.925) & 0.839 & 0.526 & 0.814 & 0.842 & 0.389 & 0.973 \\
NaiveBayes & 0.845 (0.788--0.907) & 0.795 & 0.467 & 0.814 & 0.793 & 0.327 & 0.972 \\
Neural Networks & 0.887 (0.841--0.930) & 0.808 & 0.483 & 0.814 & 0.807 & 0.343 & 0.972 \\ \hline
\end{tabular}
\label{tab: Results of the Test Set}
\end{adjustwidth}
\end{table}

\newpage

\subsection*{SHAP Analysis and Clinical Implications}

\begin{figure}[H]
    \centering
    \includegraphics[width=0.8\linewidth]{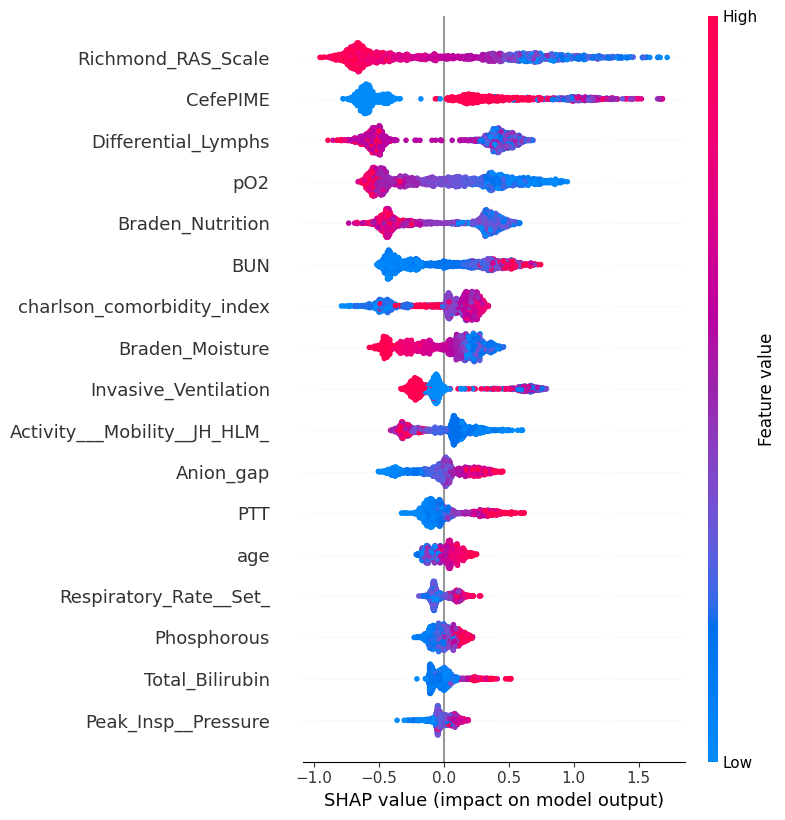}
    \caption{\textbf{SHAP summary plot showing feature contributions to predicted 30-day mortality in hypertensive ICU patients with atrial fibrillation.}}
    \label{fig:shap_summary}
\end{figure}

To further explore the contribution of individual characteristics in the CatBoost model that predicts 30-day mortality in ICU patients with coexisting hypertension and atrial fibrillation, we performed a SHAP value analysis.
SHAP is a unified framework for interpreting machine learning models by attributing the contribution of each feature to a model's output\cite{lundberg2017unified}. It is based on cooperative game theory and provides a way to fairly distribute the "credit" for a prediction across all features. SHAP values offer insights into both the magnitude and direction of each feature’s influence on the model’s prediction, making them especially useful for understanding complex models like CatBoost. By calculating SHAP values, we can better interpret how individual characteristics contribute to predictions, helping to improve model transparency and trustworthiness. Fig.~\ref{fig:shap_summary} visualizes both the magnitude and the direction of impact of each variable throughout the cohort, offering a fine-grained interpretability of the model decision behavior.

Through SHAP analysis, complemented by expert clinical input, we identified key predictors of 30-day mortality, including the Richmond-RAS Scale, CefePIME administration, invasive ventilation, and pO\textsubscript{2} levels. Additionally, lymphocyte proportion, Braden Nutrition and Mobility scores, as well as BUN and PTT values, were found to be significant predictors.

In detail, SHAP analysis revealed that the Richmond-RAS Scale was the most influential predictor of 30-day mortality. Lower values of Richmond-RAS Scale, indicative of deep sedation or reduced consciousness, are consistently associated with a marked increase in predicted mortality, aligning with the clinical understanding that reduced neurologic responsiveness correlates with worse outcomes. Similarly, the administration of CefePIME, a broad-spectrum antibiotic often used in septic or critically unstable patients, is closely linked to an increased predicted risk - especially when applied in high doses or early in the ICU stay, suggesting that antimicrobial escalation captures the underlying clinical deterioration. Low proportions of lymphocytes, often seen in systemic inflammation or sepsis, showed a clear association with elevated risk, while higher pO\textsubscript{2} levels were protective. Interestingly, this protective effect diminishes or reverses beyond certain thresholds, possibly reflecting over-oxygenation or the confounding influence of ventilatory parameters. Braden Nutrition and Mobility scores were also strongly predictive. Lower values in these domains - reflecting nutritional deficiency and immobility - were associated with higher mortality, consistent with frailty and impaired reserve in the population of hypertension and atrial fibrillation. Likewise, elevated BUN and PTT values contributed positively to risk, underscoring renal and coagulative dysfunction as key physiological stressors.

In general, this SHAP analysis confirms that the model relies on a clinically coherent and multidimensional set of predictors that range from the level of consciousness and metabolic state to comorbidities and therapeutic interventions. Such alignment between data-driven insights and medical logic reinforces the interpretability and reliability of the model in real-world ICU settings.

\subsection*{ALE Analysis and Clinical Interpretability}

\begin{figure}[H]
\begin{adjustwidth}{-2.25in}{0in}
    \centering    \includegraphics[width=0.9\linewidth]{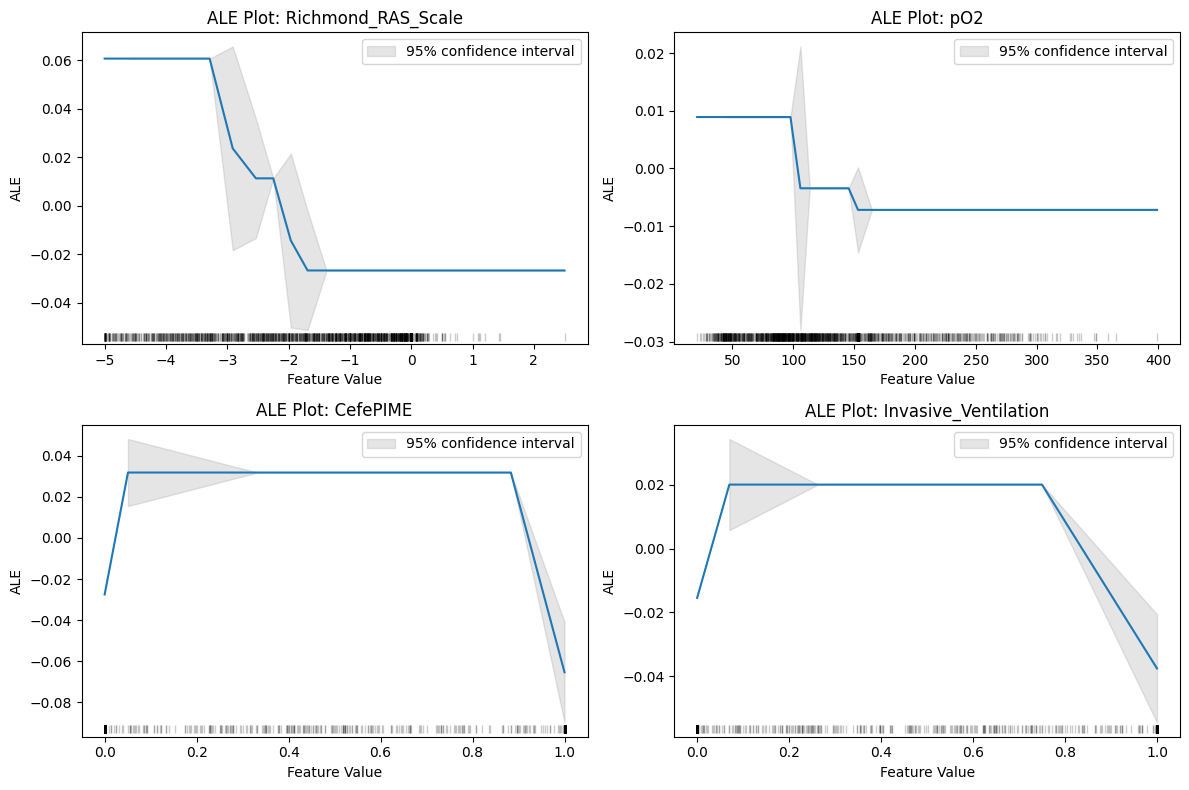}
    \caption{\textbf{ALE plots for high-impact features in ICU patients with hypertension and atrial fibrillation.}}
    \label{fig:ale_analysis_htaf}
\end{adjustwidth}
\end{figure}

To investigate the clinical reasoning embedded in the model that predicts 30-day mortality among ICU patients with coexisting hypertension and atrial fibrillation, we performed an ALE analysis on four key variables identified as having high importance of features. ALE is a model-agnostic interpretability technique that examines the impact of individual features on model predictions while accounting for interactions between features\cite{apley2020visualizing}. Unlike partial dependence plots, which assume feature independence, ALE calculates the local effect of each feature by observing how model predictions change as the feature varies, while keeping other features fixed. This approach reduces bias from correlated features and provides a more accurate interpretation of a feature’s contribution to predictions. Fig.~\ref{fig:ale_analysis_htaf} displays the ALE curves for the Richmond-RAS scale, pO\textsubscript{2}, administration of CefePIME and the use of invasive ventilation.

The ALE plot for the Richmond-RAS scale shows a strong inverse relationship between sedation depth and survival probability. Values below --2 are associated with a sharp decline in predicted survival in the model, reflecting clinical patterns where deep sedation or coma corresponds to increased severity of the disease. The monotonic and smooth curve confirms the consistency of the model in its handling of neurological status, supported by a dense sample representation throughout the range of observed values.

For pO\textsubscript{2}, the ALE curve shows a non-linear relationship: Patients with moderate oxygenation (approximately 80–120 mmHg) are associated with slightly reduced risk, while values greater than 150 mmHg show negligible marginal impact. The dip in ALE values near 100 mmHg aligns with optimal oxygenation targets in ICU protocols. The increased uncertainty at extreme values suggests reduced sample density in hyperoxic patients, which cautions interpretation in those cases.

Binary exposure to CefePIME is associated with increased predicted mortality. The ALE curve increases for values transitioning from 0 to 1, indicating that the presence of this broad-spectrum antibiotic may serve as a proxy for sepsis or high clinical concern. However, the magnitude of effect remains modest, suggesting that while predictive, it does not dominate model output. This subtle behavior adds interpretability by avoiding overreliance on treatment markers.

Invasive ventilation displays a similar binary pattern: ALE values rise for ventilated patients, again likely reflecting their increased clinical acuity. Although some statistical overlap exists, the directionality is consistent and confidence intervals are tight around the central estimate. This reinforces the model’s ability to distinguish severity-associated interventions without hard-coding clinical assumptions.

Altogether, the ALE plots illustrate the model’s nuanced capacity to encode clinically interpretable risk gradients from both physiological values and treatment proxies. The consistency and shape of each effect function, together with their alignment with the ICU practice patterns, underscore the biological plausibility of the model. These findings enhance user confidence in applying this model for triage and mortality risk stratification in complex, multimorbid ICU populations.

\subsection*{Posterior Estimation of Individual Mortality Risk}

To assess the real-world applicability of the CatBoost model for predicting 30-day mortality in ICU patients with coexisting hypertension and atrial fibrillation, we generated posterior estimates using the DREAM framework. Feature priors were sampled from empirical distributions derived from non-survivors (as detailed in Table~\ref{tab:cohort comparison results 1}), effectively simulating a critically ill, high-risk patient profile rooted in observed ICU data.

\begin{figure}[H]
    \centering
    \includegraphics[width=0.87\linewidth]{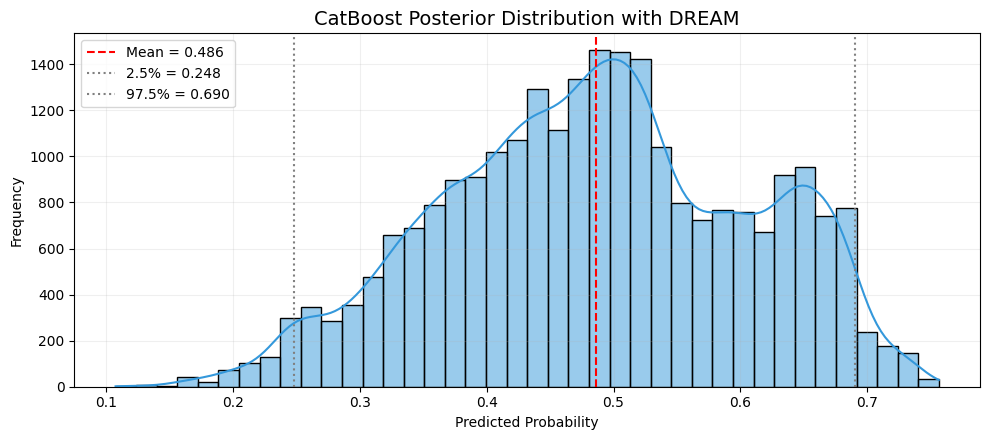}
    \caption{\textbf{DREAM analysis-posterior distribution of 30-day mortality}}
    \label{fig:uq_catboost_htaf}
\end{figure}

Fig.~\ref{fig:uq_catboost_htaf} illustrates the posterior distribution of predicted mortality for this synthetic patient. The distribution centers around a mean of 0.486, with 95\% credible intervals spanning from 0.248 to 0.690. Although the mean value approaches the probabilistic midpoint, the upper quantile reflects the model's ability to capture the mortality risk under uncertainty of the parameters, representing the plausible range for a patient in deterioration. Compared to the average event rate in the entire sample (0.196), the posterior increases risk sharply, confirming the efficacy of stratification.

The simulated input highlights several primary risk features in the ICU setting. Chief among these are a severely negative Richmond-RAS scale score ($-2.50$), reflecting deep sedation; a low pO\textsubscript{2} value (99.3 mmHg), bordering on hypoxia; and the presence of invasive ventilation and CefePIME administration, indicating both the need for mechanical respiratory support and the use of broad-spectrum antibiotics for suspected or confirmed infection. These categorical variables—specifically invasive ventilation and CefePIME—directly shift the model’s risk estimation toward higher values, underscoring their strong association with adverse outcomes in acute ICU care.

In addition to these primary predictors, other features further reinforce the model’s assessment of physiological stress and mortality risk. Elevated BUN and prolonged PTT indicate renal dysfunction and coagulopathy, respectively. Impaired mobility (JH-HLM = 2.00) and a high Charlson comorbidity index (4.59) reflect both functional decline and substantial chronic disease burden. Continuous variables such as Braden Nutrition score and anion gap provide further refinement to the risk estimation by capturing subtle aspects of metabolic status and frailty. Together, these features illustrate how the model integrates acute interventions, physiological derangements, and baseline vulnerability to inform prognostic assessment.

The DREAM-based posterior output provides a probabilistic interpretation of model inference, enabling clinicians to understand not only the point estimate of risk but also the uncertainty surrounding it. In high-stakes ICU settings, this uncertainty quantification supports more informed and transparent decision-making. Rather than offering a fixed threshold, the model communicates the likelihood of death as a spectrum—allowing for individualized clinical reasoning grounded in data-driven simulations.

\section*{Discussion}
\subsection*{Summary of Existing Model Compilation}

In this study, we proposed a clinically interpretable and practically deployable machine learning framework aimed at predicting 30-day in-hospital mortality in ICU patients presenting with both hypertension and AF. This dual diagnosis is common among critically ill individuals, yet few prediction models have focused explicitly on this high-risk subgroup. 

This study utilized the MIMIC-IV database, one of the largest publicly available critical care datasets, to extract a well-defined cohort of 1,301 patients with concurrent hypertension and AF. A stringent and systematic pipeline was applied, excluding patients with terminal malignancies and considering only their first ICU admission. This approach ensured that the model was focused on early-phase, generalizable ICU cases.

This study incorporated a comprehensive set of features, including chart events, laboratory events, procedure events, drug administration, and demographic and comorbidity data. An initial pool of nearly 300 features was extracted from data within the first 24 hours of ICU admission for this population. Utilizing MI ranking techniques in conjunction with expert knowledge, we identified 17 critical features for predicting 30-day mortality in this population. These features included chart events such as BUN, Richmond-RAS Scale, PTT, phosphorus, total bilirubin, anion gap, differential lymphocytes, Braden Nutrition, Braden Moisture, respiratory rate (set), activity/mobility (JH-HLM), and peak inspiratory pressure; laboratory events, specifically pO\textsubscript{2}; procedure events, including invasive ventilation; drug administration, notably CefePIME; and demographic/comorbidity factors such as age and the Charlson Comorbidity Index.

This study applied six machine learning algorithms, with CatBoost serving as the primary model and LightGBM, XGBoost, Random Forest, Logistic Regression, and SVM serving as baseline models. To ensure model robustness, stratified cross-validation was employed, and hyperparameters were optimized through grid search. CatBoost emerged as the top performer, achieving an AUROC of 0.889(95\%CI:0.840-0.924) on the test set. Performance across other evaluation metrics was also notable, demonstrating robustness in the model’s predictive capabilities.

In addition to evaluating predictive performance, this study placed a strong emphasis on model interpretability and clinical usability. Specifically, a combination of ablation analysis, SHAP, ALE, and DREAM analyses was employed to elucidate model predictions at both the global and local levels. This multi-faceted interpretability framework enabled transparent identification of key mortality-associated features, including the Richmond-RAS Scale, pO\textsubscript{2}, invasive ventilation, and CefePIME administration. Importantly, this approach provides a novel perspective and methodology for understanding and communicating the determinants of short-term mortality risk in the specific ICU population.

By combining methodological rigor with clinical relevance, our model contributes not only a high-performing predictive tool but also an interpretable, implementable solution for ICU decision-making.

\subsection*{Comparison with Prior Studies}

A number of previous studies have examined the intersection of atrial fibrillation and hypertension using traditional statistical approaches; however, most have focused on general or outpatient populations and have aimed to identify long-term cardiovascular risks rather than short-term mortality in ICU settings. 

For instance, Li et al.\cite{li2024nomogram} developed a nomogram to predict 1-year AF risk in hypertensive patients based on logistic regression. Their study emphasized biomarkers such as HbA1c and lipoprotein(a), reporting an AUROC of 0.793. While useful for preventive cardiology, the study’s population, time horizon, and outcome differ substantially from ours. It is not designed to guide acute decisions or critical care triage.

Wang et al.\cite{wang2023prevalence} used Cox proportional hazards models to study the association between elevated diastolic blood pressure and new-onset AF, as well as mortality in hypertensive patients. Although the study highlighted important prognostic associations, it lacked predictive modeling, explainability, and ICU specificity. Similarly, other works using regression analysis have identified hypertension as a risk factor for AF incidence or stroke, but few have modeled its interaction with AF in the context of critical care.

Unlike these studies, our work fills a clinically meaningful gap by developing a real-time, interpretable model specifically for ICU patients with AF and hypertension. This population is of particular interest because these conditions often co-exist yet may require competing treatment strategies—e.g., rate control versus hemodynamic stabilization—which complicates clinical management. Our model, trained on a relevant patient population and constrained to early-phase data, is designed for immediate clinical action, not retrospective risk scoring.

Compared to regression-based tools, our model captures non-linear relationships and complex interactions, as demonstrated through a combination of an ablation study, SHAP, ALE and DREAM explanations. Importantly, our model does not rely on long-term lab trends or specialized imaging and can be implemented directly within other EHR systems with minimal overhead.

Moreover, few prior studies offer models with clinical transparency suitable for bedside use. While some ML-based ICU studies report high AUROC values, they often fall short in explainability and patient-specific interpretability\cite{nemati2018interpretable,zhang2022prediction,wang2023machine}. Our interpretable methods allow ICU clinicians to understand not only who is at risk but also why they are at risk, enabling informed conversations about care escalation or withdrawal.

In summary, while existing literature has addressed components of the AF-hypertension relationship, none have combined short-term mortality prediction, interpretable ML modeling, ICU specificity, and real-time deployment potential as comprehensively as our study. This positions our work as a valuable and practical contribution to precision critical care.

\subsection*{Limitations and Future Work}

Despite its strengths, this study has several limitations. Firstly, it is a single-center retrospective study using data solely from MIMIC-IV, which may limit its external validity. Although MIMIC-IV includes diverse ICU populations and remains one of the most widely used databases in critical care research, prospective multicenter validation will be essential to ensure generalizability. Secondly, some clinically informative features—such as echocardiographic parameters, medication dosages, and fluid balance—were unavailable or incompletely recorded, and thus not included in the model. Thirdly, the model uses only static features from the first 24 hours of ICU admission. While this improves feasibility for early prediction, it excludes dynamic trends that may offer additional prognostic information. Future versions may incorporate sequential models such as LSTM or attention-based architectures to capture temporal patterns. Furthermore, while we emphasized interpretability through SHAP and ALE, real-world deployment will require the development of clinician-facing interfaces, integration with EHRs, and usability testing to support adoption. Lastly, our modeling approach and interpretability framework can be adapted to other critical comorbidity pairs beyond hypertension and AF. Future studies may explore more diseases based on our framework, to contribute more to the clinical decision-making.

\section*{Conclusion}

In this study, we developed and validated a clinically interpretable machine learning framework to predict 30-day in-hospital mortality in ICU patients with coexisting hypertension and atrial fibrillation. 

Using a curated cohort of 1301 patients with Hypertension and AF from the MIMIC-IV database, a total of 17 clinically relevant variables were selected for model development, reflecting chart events, laboratory events, procedure events, drug administration, and demographic and comorbidity data within the first 24 hours of ICU admission. 

The proposed CatBoost model demonstrated strong predictive performance, achieving an AUROC of 0.889 (95\% CI: 0.840–0.924), with an accuracy of 0.831, F1-score of 0.522, sensitivity of 0.837, specificity of 0.830, PPV of 0.379, and NPV of 0.976, underscoring its robustness in predicting 30-day mortality in ICU patients with coexisting hypertension and atrial fibrillation.

Global and local explanation methods, including an ablation study, SHAP, ALE, and DREAM analysis, facilitated the transparent identification of key predictors such as the Richmond-RAS Scale, pO\textsubscript{2}, CefePIME, and Invasive Ventilation. By identifying and quantifying the contributions of these key predictors, the study offers valuable perspectives and evidence for clinical decision-making, empowering healthcare professionals with more comprehensive tools for early intervention and personalized care. The integration of such methods ensures that the model's predictions are not only accurate but also clinically actionable, enriching the decision-making process with greater transparency and accountability.

This study emphasizes the clinical value of interpretable machine learning for real-time risk assessment in high-risk ICU populations. By targeting the intersection of hypertension and atrial fibrillation, our model proves to be an effective tool for early intervention and personalized care decisions. Future directions include external validation with multi-center ICU datasets, incorporating longitudinal data to track patient trajectories, and applying this approach to other diseases using the developed pipeline.

\nolinenumbers

%
%
%
\bibliography{referrence}

\end{document}